\documentclass[aps,superscriptaddress,showpacs]{revtex4}
\usepackage{graphicx}
\usepackage{subfigure}
\usepackage{bm}
\begin{document}

\title{Statistics and geometry of passive scalars in turbulence}
\author{J\"org Schumacher}
\affiliation{Fachbereich Physik, Philipps-Universit\"at, Renthof
6, D-35032 Marburg, Germany}
\author{Katepalli R. Sreenivasan}
\affiliation{International Centre for Theoretical Physics, 34014 Trieste, Italy}
\date{\today}

\begin{abstract}

We present direct numerical simulations (DNS) of the mixing of the
passive scalar at modest Reynolds numbers ($10 \le R_\lambda \le
42$) and Schmidt numbers larger than unity ($2 \le Sc \le 32$).
The simulations resolve below the Batchelor scale up to a factor
of four. The advecting turbulence is homogeneous and isotropic,
and maintained stationary by stochastic forcing at low
wavenumbers. The passive scalar is rendered stationary by a mean
scalar gradient in one direction. The relation between geometrical
and statistical properties of scalar field and its gradients is
examined. The Reynolds numbers and Schmidt numbers are not large
enough for either the Kolmogorov scaling or the Batchelor scaling
to develop and, not surprisingly, we find no fractal scaling of
scalar level sets, or isosurfaces, in the intermediate viscous
range. The area-to-volume ratio of isosurfaces reflects the nearly
Gaussian statistics of the scalar fluctuations. The scalar flux
across the isosurfaces, which is determined by the conditional
probability density function (PDF) of the scalar gradient
magnitude, has a stretched exponential distribution towards the tails. The
PDF of the scalar dissipation departs distinctly, for both small
and large amplitudes, from the lognormal distribution for all
cases considered. The joint statistics of the scalar and its
dissipation rate, and the mean conditional moment of the scalar
dissipation, are studied as well. We examine the effects of
coarse-graining on the probability density to simulate the effects
of poor probe-resolution in measurements.

\end{abstract}
\maketitle

\section{Introduction}
The mixing of passive scalars in the presence of turbulent motion
is a subject of great theoretical and practical interest. Examples
of its ubiquitous applications arise in reacting flows and
combustion, mixing of salt and plankton in oceans and of
pollutants in the atmosphere, as well as mixing of biological
substances. Much experimental effort has thus been invested on
mapping scalar fields in three dimensions, usually by
reconstructing them from several planar cuts taken in quick
succession in time; see Refs.~1 to 6 for work on turbulent liquid
or air jets and Refs.~7 and 8 for work on diffusion flames. The
Schmidt number $Sc=\nu/\kappa$ where $\nu$ is the kinematic
viscosity of the advecting fluid and $\kappa$ is the scalar
diffusivity, plays an important role in determining the nature of
mixing \cite{batchelor}. For the following, our focus will be on
Schmidt numbers larger than unity.

The finest scales of the scalar are generally thought to be of the
order of the Batchelor scale defined as $\eta_B=\eta/\sqrt{Sc}$
where $\eta$ is the Kolmogorov scale given by $\eta = \nu^{3/4}/
\langle \epsilon \rangle$; $\epsilon$ is the local energy
dissipation rate of turbulence, and the angular brackets indicate
a suitable average. For a given flow, the Batchelor scale becomes
smaller for larger $Sc$. Thus, the required resolution for large
$Sc$, common in liquid flows, is hard to attain experimentally
\cite{bilger04}.

With modern computers, it is possible to resolve scalar mixing in
turbulence at high Schmidt numbers
\cite{yeung01,yeung02,brethouwer03,schu03,watanabe04}. Schumacher
{\it et al.} \cite{schu05} showed that it is sometimes necessary
to resolve scales that are finer than $\eta_B$. Such fine
resolutions are desired, for instance, for obtaining reliable
results on the extreme fluctuations (both small and large) of
scalar gradients and scalar dissipation rate. We resolve scales
finer than the Batchelor scale by a factor of four. This fine
resolution limits the ranges of $Sc$ and $R_\lambda$ examined,
where $R_\lambda$ is the Taylor microscale Reynolds number.
Nevertheless, we can study trends with $Sc$ and $R_\lambda$ in
limited ranges of these parameters, with full three-dimensional
field resolved extremely well. In particular, we consider
relations between geometric and statistical properties of passive
scalars and their spatial variations with respect to $Sc$ for a
given turbulent flow, as well as those with the Reynolds number
for a given $Sc$.

The paper is organized as follows. Section II briefly reports the
numerical details and parameters. We consider the box-counting
properties and area-to-volume ratio of the passive scalar isosurfaces in Sec.\
III, while Sec.\ IV considers the properties of the instantaneous
scalar flux across isosurfaces. Section V presents an analysis of
the statistics of scalar dissipation including conditional
statistics. Conclusions and summary are presented in the final
section.

\section{Numerical Simulations}
We solve simultaneously the Navier-Stokes equations for an
incompressible flow ${\bf u}({\bf x},t)$ and the
advection-diffusion equation for the passive scalar field
$\theta({\bf x},t)$ in box-type three-dimensional turbulence with
the classical pseudo-spectral method that uses fast Fourier
transformations and a 2/3 de-aliasing. The boundary conditions for the
velocity and scalar field are both periodic. The simulation domain
has a length of $2\pi$ in each coordinate direction and is
resolved by an equidistant grid with $N^3$ points. The equations
are
\begin{eqnarray}
\label{nseq}
\frac{\partial{\bf u}}{\partial t}+({\bf u}\cdot{\bf \nabla}){\bf u}
&=&-{\bf \nabla} p+\nu {\bf \nabla}^2{\bf u}+{\bf f}\,,\\
\label{ceq}
{\bf \nabla}\cdot{\bf u}&=&0\,,\\
\frac{\partial \theta}{\partial t}+({\bf u}\cdot{\bf \nabla}) \theta
&=&\kappa {\bf \nabla}^2\theta - u_y G\,.
\label{pseq}
\end{eqnarray}
Here, $p({\bf x},t)$ is the kinematic pressure field. The
turbulence is sustained by a random force density ${\bf f}({\bf
x},t)$ (see Vedula {\it et al.} \cite{yeung01} and Schumacher
\cite{schu04} for details) and the scalar gradient is constant in
the $y$ direction, i.e. ${\bf G}=(0,G,0)$ with $G=1$; $u_y$ is the
turbulent velocity component in the direction $y$. The equations
are integrated in time by a second-order predictor-corrector
scheme. Parameters of the simulations and some statistical results
are listed in Table 1. The spectral resolution exceeds the usual
criterion of $k_{max} \eta_B\ge 1.5$ by a factor of about 4. A
detailed comparison between these high-resolution data and those
with the nominal resolution employed in most DNS studies was
presented in Schumacher {\it et al.} \cite{schu05}. Their
conclusion was that super-fine resolution would be needed to
capture specific features of the scalar dissipation rate
\begin{equation}
\epsilon_{\theta}({\bf x},t)=\kappa ({\bf\nabla} \theta ({\bf
x},t))^2\,,
\label{scalardiss}
\end{equation}
such as the sheet-like structures in which the extreme events are
found. We shall exploit the present fine resolution for just such
purposes.
\begin{table}
\begin{center}
\begin{tabular}{lccccccc}
\hline\hline
\multicolumn{1}{l}{Run No.} & \multicolumn{3}{c}{1} &\multicolumn{3}{c}{2} &\multicolumn{1}{c}{3} \\
\hline
\multicolumn{1}{l}{$N$} & \multicolumn{3}{c}{512}& \multicolumn{3}{c}{1024}&\multicolumn{1}{c}{1024}  \\
\multicolumn{1}{l}{$\nu$}  & \multicolumn{3}{c}{0.0333}& \multicolumn{3}{c}{0.0133}& \multicolumn{1}{c}{0.005} \\
\multicolumn{1}{l}{$\langle\epsilon\rangle$} & \multicolumn{3}{c}{0.1}& \multicolumn{3}{c}{0.1}&\multicolumn{1}{c}{0.1}  \\
\multicolumn{1}{l}{$u_{rms}$} & \multicolumn{3}{c}{0.394} & \multicolumn{3}{c}{0.483}&\multicolumn{1}{c}{0.486}  \\
\multicolumn{1}{l}{$k_m \eta $} &  \multicolumn{3}{c}{33.56}&  \multicolumn{3}{c}{33.56}&\multicolumn{1}{c}{15.93}  \\
\multicolumn{1}{l}{$R_{\lambda}$} &\multicolumn{3}{c}{10} &
                                \multicolumn{3}{c}{24} & \multicolumn{1}{c}{42} \\
\multicolumn{1}{l}{$L$} &   \multicolumn{3}{c}{1.018}&   \multicolumn{3}{c}{0.920}&\multicolumn{1}{c}{0.758}  \\
\multicolumn{1}{l}{$T_{av}/T_E$} & \multicolumn{3}{c}{9.1}& \multicolumn{3}{c}{1.2}&\multicolumn{1}{c}{1.1}  \\
$Sc$                     & 2 & 8 & 32 & 2 & 8 & 32 & 32\\
$\theta_{rms}$   & 1.390 & 2.012 & 2.572 & 1.269 & 1.604 & 1.917 & 1.548 \\
$\langle\epsilon_{\theta}\rangle$ & 0.330 & 0.391 & 0.420 & 0.307 & 0.345 & 0.372 & 0.306\\
$k_m \eta_B$        & 23.68 & 11.84 & 5.92 & 23.68 & 11.84 & 5.92 & 2.82 \\
$L_{\theta}$  & 0.707 & 0.587 & 0.477 & 0.562 & 0.447 & 0.357 & 0.289 \\
\hline\hline
\end{tabular}
\label{tab1} \caption{List of parameters for the present DNS runs.
$N$ is the number of grid points on each side of the computational
cube, $\nu$ is the kinematic viscosity, $\langle\epsilon\rangle$
is the mean energy dissipation rate, $u_{rms} \equiv \langle u_x^2
\rangle^{1/2}$ is the root-mean-square velocity fluctuation in $x$
direction, $k_m=\sqrt{2}N/3$ is the maximum wavenumber allowed in
the simulations,
$R_\lambda=\sqrt{15/(\nu\langle\epsilon\rangle)}\,\langle u_x^2
\rangle$ is the Taylor microscale Reynolds number, $L$ is the
integral scale of turbulence. Its magnitude is to be compared with
the linear dimension of the box, which is $2 \pi$ in all cases.
$T_{av}$ is the statistical averaging time, to be compared here
with the large-eddy turnover time defined as $T_E = \langle
{\bf u}^2\rangle/2\langle\epsilon\rangle$; $Sc$ is the Schmidt number,
$\theta_{rms}\equiv \langle\theta^2\rangle^{1/2}$ is the
root-mean-square passive scalar fluctuation,
$\langle\epsilon_{\theta}\rangle$ is the mean scalar dissipation
rate and $L_{\theta}$ is the integral scale of the passive
scalar. Both integral scales are calculated as in Ref. 44.}
\end{center}
\end{table}

\begin{figure}
\centerline{\includegraphics[angle=0,scale=1.5,draft=false]{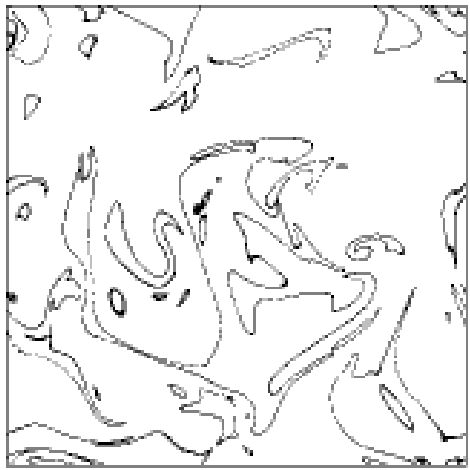}}
\caption{Two-dimensional instantaneous slice
through the fluctuating passive scalar field $\theta({\bf x},t)$
for the level of $0.8\,\theta_{rms}$ (i.e., $X = 0.8$). The
side-length of the box (Run 2) is about 90$\eta$ and resolved
with $N=1024$ in each coordinate direction. $R_\lambda = 24$, $Sc = 32$.}
\end{figure}
\begin{figure}
\centerline{\includegraphics[angle=0,scale=0.8,draft=false]{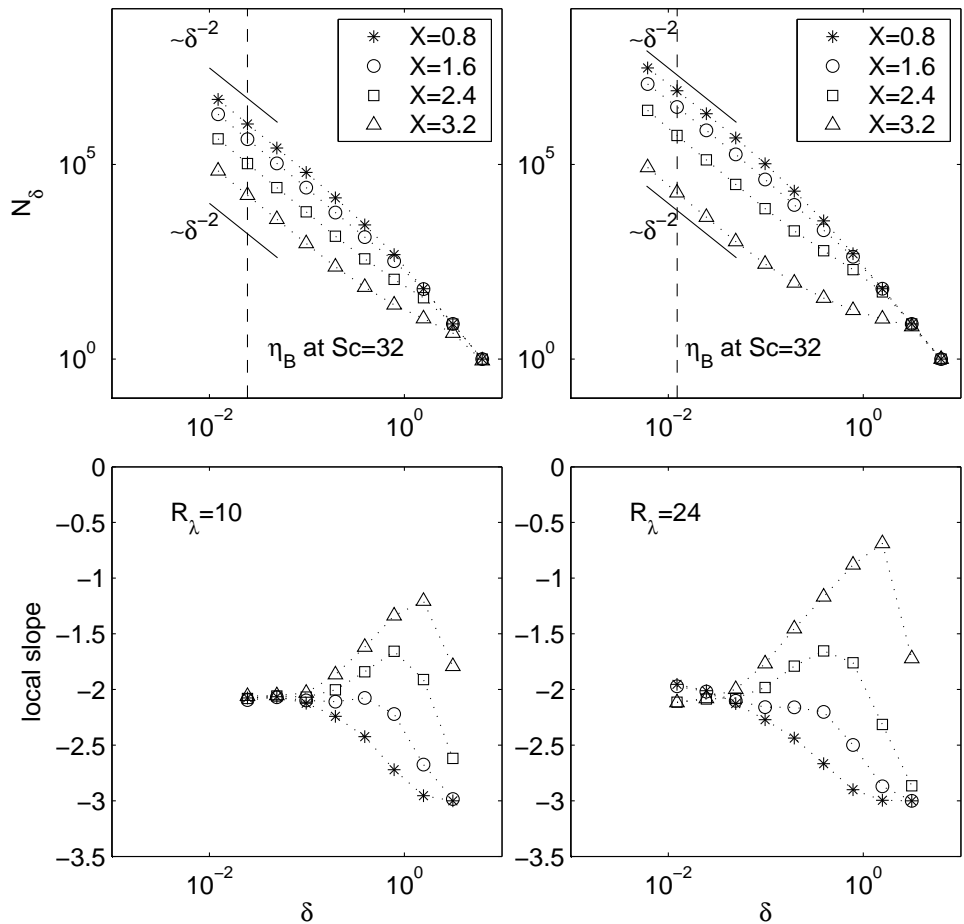}}
\caption{Three-dimensional box-counting data for
scalar level sets $L_{X}$ as defined in (\ref{deflevelset}) for
$Sc=32$ and for $R_{\lambda}=10$ (left column) and
$R_{\lambda}=24$ (right column). The values of $X$ are: asterisks,
0.8; circles, 1.6; squares, 2.4; triangles, 3.2. The vertical
dashed lines in the upper panels show the corresponding values of
$\eta_B$. Short solid lines mark the algebraic slope with
$\delta^{-2}$ in the log-log plot. All the isosurfaces have a
region with a common exponent of 2 at the smallest scales. The
lower panels show the local slope
$\mbox{d}\log(N_{\delta})/\mbox{d}\log(\delta)$ as a function of
$\delta$. Symbols are as in the upper figures.}
\end{figure}

\section{Scalar isosurfaces}
\subsection{Box-counting dimension}
It is well-known that, for Schmidt numbers exceeding unity, the
passive scalar is filamented in structure and mixed by smooth flow
in a continual stretch-twist-fold scenario \cite{Ott}. An open
point is whether the scalar contours accompanying such dynamics,
when diffusion and external driving are included, generate fractal
sets. Neither the inertial-convective range nor the
viscous-convective Batchelor range obtains for the present
moderate values of $R_\lambda$ and $Sc$, so it would be surprising
if the fractal scaling is found. Nevertheless, the issue is worth
a brief examination as a prelude to the study of area-to-volume
ratio of scalar interfaces that follows in Sec.\ III\ B. If the
surface is indeed a fractal, it is easy to write down the surface
area in terms of its dimension and the cut-off scales terminating
the scaling behavior \cite{sreeni}.

Let us define the scalar level sets as
\begin{equation}
L_{X}=\{{\bf x}: X-\delta X\le\theta({\bf x},t)/\theta_{rms}\le
X+\delta X\}\,, \label{deflevelset}
\end{equation}
where $X$ is the ratio of the chosen amplitude threshold to the
root-mean-square value of the scalar, $\theta_{rms}$.
Operationally, $\delta X$ is taken to be $2.5\%$ of $\theta_{rms}$ for each
$(R_{\lambda},Sc)$. Figure~1 shows a level set of the scalar field
around a value of $0.8 \theta_{rms}$ for $Sc = 32$ and
$R_{\lambda}=24$. Similar pictures with different thresholds show
that levels of high intensity are embedded within iso-levels of
lower intensity. The iso-levels become increasingly
disconnected with increasing intensity and become isolated islands
embedded within an iso-contour of lower magnitude. This
observation is consistent with mixing studies in a jet by
Villermaux and Innocenti \cite{villermaux99}. In the box-counting
procedure described, for example, in Falconer \cite{Falconer}, one
counts the number of cubes, $N_{\delta}$, of side-length $\delta$
that cover the level set $L_X$ completely. This procedure is
repeated for various values of $\delta$. The box-counting
dimension $D$ of a (fractal) level set is defined then as
\begin{equation}
D=-\lim_{\delta\to 0}\;\frac{\log N_{\delta}}{\log \delta}\,.
\label{defdimension}
\end{equation}
In practice, it is clear that one cannot apply the limit
$\delta\to 0$ and one seeks an algebraic scaling,
$N_{\delta}\sim\delta^{-D}$, in a finite range.

The upper two panels of Fig.\ 2 show the log-log plots resulting
from box-counting for data sets of Runs 1 and 2 at $Sc=32$. A
constant slope with exponent $D$ would give the box-counting
dimension of the scalar level set. However, the local slope given
by
\begin{equation}
D(\delta)=-\frac{\mbox{d}\log N_{\delta}}{\mbox{d}\log \delta}\,
\label{deflocalslope}
\end{equation}
varies continuously, as shown in the lower panels of Fig.~2.
Though our simulations cover three orders of magnitude of scales,
it is clear that no power-law scaling exists for reasons mentioned
at the beginning of this section. The smallest few values of
$\delta$ give a local slope of 2, for all $X$, which means that
the level sets at those scales become smooth sheets. We interpret
(the slightly) smaller slopes for intermediate scales of $\delta$
as arising from the disconnected shape of the iso-level sets.
Indeed, this patchiness increases with the level $X$ thus giving a
smaller magnitude of the local slope.

The conclusion from this study is that fractal behavior of the
isolevel sets is not obtained in the range of $R_\lambda$ and $Sc$
considered here. We want to emphasize, however, that this finding
has no bearing on the past experimental work as reported in
Refs.~3, 4, 19 and 21. The
differences such as may exist in these experimental studies
require a separate discussion which lies outside the scope of this
paper. We expect that the fractal property in the
inertial-convective range requires high Reynolds numbers while
that in the viscous-convective range requires high values of
Schmidt {\it and} Reynolds numbers.
\begin{figure}
\centerline{\includegraphics[angle=0,scale=0.7,draft=false]{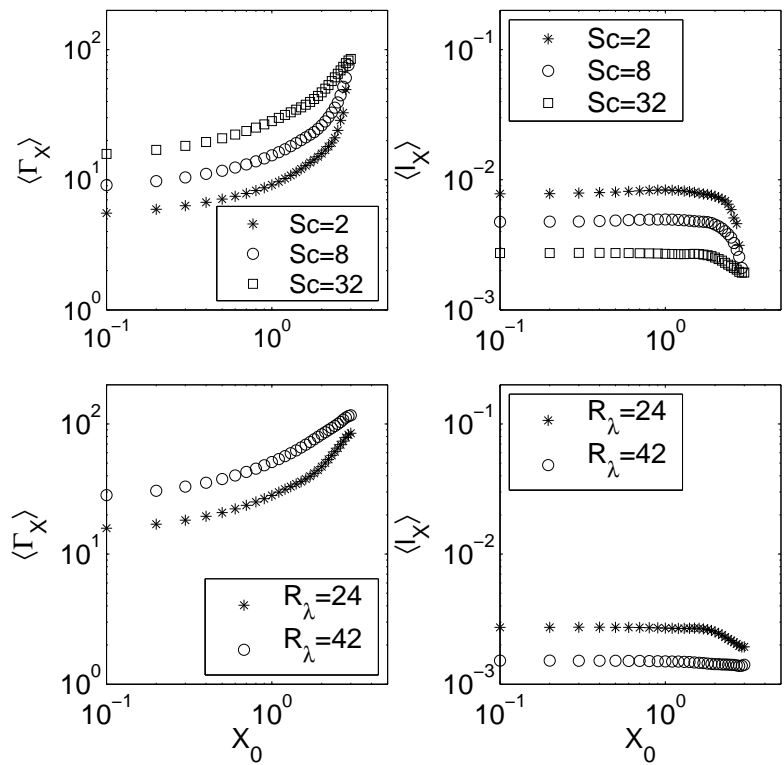}}
\caption{Left column: mean of the area-to-volume
ratio $\langle\Gamma_X\rangle$ as a function of the isolevel $X$
in a log-log plot. Upper panel shows data for Run 2 at three
different Schmidt numbers. Lower panel shows the quantity at
$Sc=32$ for Runs 2 and 3. Right column: the corresponding mean
thickness scale $\langle\ell_X\rangle$ as derived
from Eq.~(\ref{areavol2}). }
\end{figure}
\begin{figure}
\centerline{\includegraphics[angle=0,scale=1,draft=false]{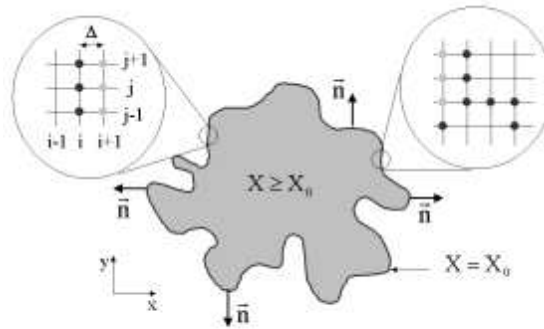}}
\caption{Sketch of the scalar gradient flux
calculation across an passive scalar isosurface $X=X_0$. The illustration is for a
two-dimensional case (but the data presented in the next figure
are for the surface in three dimensions). Black points lie on the
boundary whereas the gray points are the interior of
$\tilde{L}_X$. The inner points are the volume that is used for $\Gamma_X$
(see Eq.~(\ref{areavol1})).}
\end{figure}

\subsection{Area-to-volume ratio}
Since we do not have fractal scaling for the ranges of parameters
considered here, it is necessary to compute the area of
isosurfaces directly. In reacting flows, a quantity of some
importance is the ratio of the isosurface area content to the
volume which encloses it \cite{pope88}. This ratio is numerically
calculated as follows. We first define the set
\begin{equation}
\tilde{L}_{X}=\{{\bf x}: \theta({\bf x},t)/\theta_{rms}\ge X\}\,.
\label{deflevelset1}
\end{equation}
The set is computed for each snapshot separately and the results
are averaged with respect to time subsequently. We mark all grid
points that belong to $\tilde{L}_X$ by an indicator function
$I_{ijk}\equiv 1$. For all other points, $I_{ijk}\equiv 0$. Here
$i$, $j$, and $k$ denote the indices of the grid vertices running
from 1 to $N$. All boundary points of the set $L_X$ have a number
of nearest neighbors, $N_{ijk}$, that ranges from 1 to 5; the
inner points of $\tilde{L}_X$ have 6 nearest neighbors. The number
of inner points is denoted by $\overline{N}_{ijk}$. The
area-to-volume ratio $\Gamma_X$ of the set $\tilde{L}_X$ is given
by
\begin{equation}
\Gamma_X=\frac{A_X}{V_X}=\frac{1}{\overline{N}_{i,j,k}\Delta x}\sum_{(i,j,k)\in\tilde{L}_X}\,(6-N_{ijk})\,,
\label{areavol1}
\end{equation}
with grid spacing $\Delta x$, with $\overline{N}_{i,j,k}\ne 0$.
The left column of Fig.~3 shows the average of the area-to-volume
ratio, with $\langle\Gamma_X\rangle$ taken over several snapshots.
Three trends are clear: first, the ratio grows with increasing
$X$, which indicates a higher degree of disconnection of the
isosurfaces. Second, the ratio is larger for larger Schmidt
numbers. This is because mixing then occurs over ever finer scales
whose contours are stretched and folded by the flow. Finally, a
larger Reynolds number enables the scalar to be stirred more
efficiently on all scales, yielding larger area-to-volume ratios
with increasing Reynolds number.

We relate this geometric measure to statistical properties of the
passive scalar fluctuations following ideas that were used in the
slightly different context of interfacial dynamics for a turbulent
jet \cite{aguirre04}. As is shown later (see Sec.\ V\ B), the scalar
fluctuations are close to the Gaussian probability density function,
$p(X)=1/\sqrt{2\pi}\,\exp(-X^2/2)$, with a slight sub-Gaussian
behavior in the far tails. The average of the ratio
$\langle\Gamma_X\rangle$ should therefore be related to $p(X)$. For
$X=X_0$ we set
\begin{equation}
\langle\Gamma_{X_0}\rangle\,\langle\ell_{X_0}\rangle \simeq
\frac{\exp(-X_0^2/2)\,\mbox{d}X_0}{\int_{X_0}^{\infty}\,\exp(-y^2/2)\,\mbox{d}y}=
\frac{\exp(-X_0^2/2)\,\mbox{d}X_0}{\sqrt{\pi/2}\left[1-\mbox{erf}(X_0/\sqrt{2})\right]}\,,
\label{areavol2}
\end{equation}
where $\mbox{erf}(x) = (2/\sqrt{\pi})\int_0^x\,\exp(-y^2)\,
\mbox{d}y$ is the error function. For the following, we discuss
results for values $X_0>0$, but everything holds for $X_0<0$ just
the same way. The additional scale $\langle\ell_{X_0}\rangle$ on the
left hand side of Eq.\ (10) has to appear for dimensional reasons
since the right hand side is dimensionless.

The scale $\langle\ell_{X_0}\rangle$ is related to the
surface-to-volume ratio $\Gamma$, and is small when the surface-to-volume ratio
is large. Since $\Gamma$ becomes larger for ``rougher" surfaces, the
scale is a characteristic measure of the transition between smooth
and rough isosurfaces, perhaps describing the average connectedness and
regularity of the isosurface. We may thus call it the transition
scale. It is a global average measure as well as $\Gamma$.  Its trends with $Sc$ and 
$R_{\lambda}$ are shown in the right
column of Fig.~3. The straight horizontal parts of $\langle\ell_{X}\rangle$ for $X\lesssim 2$
at fixed $Sc$ and $R_{\lambda}$
are the ``geometric fingerprints" of the Gaussianity of the scalar
fluctuations. Finally, we
want to note that the relation (\ref{areavol2}) holds only for
isosurfaces that are not fractal.
\begin{figure}
\centerline{\includegraphics[angle=0,scale=0.8,draft=false]{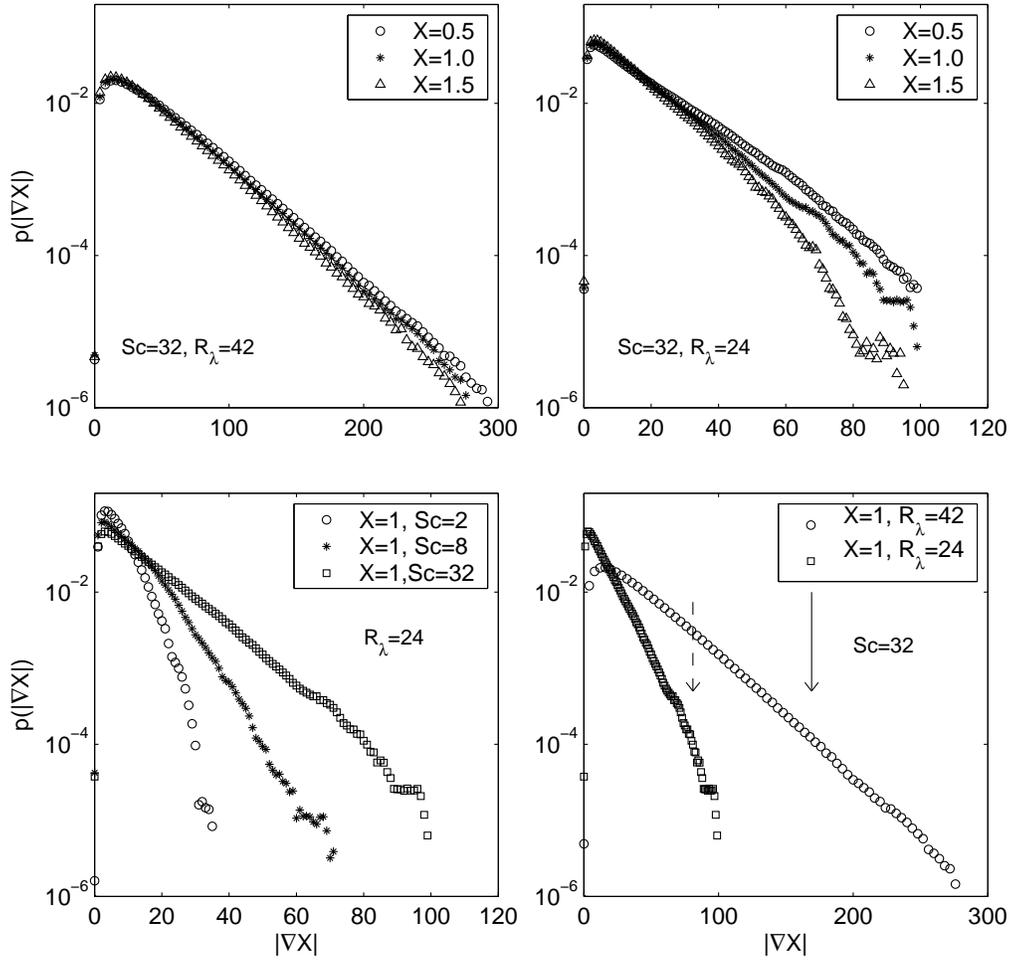}}
\caption{The probability density function of the
magnitude of scalar gradient, $|{\bf\nabla}X|$, taken at
iso-contour levels $X=X_0$. Upper left:  The data are for Run 3,
$Sc=32$, for three different levels: $X=0.5$ (open circle),
$X=1.0$ (asterisk), $X=1.5$ (triangle). Upper right: The data are
for Run 2, $Sc=32$, for three different levels: $X=0.5$ (open
circle), $X=1.0$ (asterisk), and $X=1.5$ (triangle). Lower left:
Schmidt number dependence of the flux. Data are for Run 2 at
$X=1$: $Sc=2$ (circle), $Sc=8$ (asterisk), and $Sc=32$ (square).
Lower right: Reynolds number dependence of the flux for $Sc=32$
and for $X=1$, with $R_{\lambda}=24$, Run 2 (square) and
$R_{\lambda}=42$, Run 3 (circle). The vertical arrows (dashed for
Run 2 and solid for Run 3) indicate $X/\eta_B$ with $X=1$.}
\end{figure}

\section{Flux across scalar isosurfaces}
One reason for the interest in the area of an isosurface is the
flux across it. For example, in non-premixed turbulent combustion
the isosurface of the stochiometric mixture fraction plays an
important role and the variation of gradients across it are
important for closure models \cite{pope88,bilger04,verwisch98}.
This can be calculated by integrating the flux across all
infinitesimal elements of the surface. Since the flux is
proportional to the product of the area and the gradient of the
concentration across it, it is necessary to know the gradient
across each infinitesimal element of the surface. The gradient can
itself be a highly fluctuating variable.

The present data allow the computation of gradients and fluxes
across infinitesimal elements of isosurfaces. The (differential)
flux across an infinitesimal area element $\mbox{d}A$ of an
isosurface is given by
\begin{equation}
\mbox{d}\Phi(X_0)=-{\bf\nabla}X|_{X=X_0} {\bf\cdot n}\,\mbox{d}A\,,
\label{diffflux}
\end{equation}
where ${\bf n}$ is the outward unit normal (see Fig.\ 4 for
$X_0>0$).

The probability density function (PDF) of $\mbox{d}\Phi(X_0)$ is
then fully determined by the PDF of scalar gradient magnitude at
the isolevel $X=X_0$
\begin{equation}
p(\mbox{d}\Phi(X_0))=p(-|{\bf\nabla}X|_{X=X_0}\,\mbox{d}A)\,.
\label{diffflux}
\end{equation}
Thus, the quantity of interest is a conditional PDF
$p(|{\bf\nabla}X|\, |X=X_0)$ which is calculated as follows.
First, as in the preceding section, we mark all the grid points
for which $\theta({\bf x},t)\ge X\theta_{rms}$ as {\em active}
points; their grid sites are marked with $I_{i,j,k}\equiv 1$; for
all other points, $I_{i,j,k}\equiv 0$. Gradients are calculated on
$\partial \tilde{L}_X$ afterwards. Sparser grid resolutions would
require a trilinear interpolation between the grid cells
\cite{Kollmann1997}.

The result is plotted in Fig.~5. A stretched exponential behavior is
observed for a large range of the PDF, with the tails decaying
more rapidly for larger $X$. The PDFs suggests also that there is
a considerable local variation of the flux across interfaces $X=X_0$.
The fluctuations increase significantly with increase in $Sc$ as
well as $R_\lambda$. For both Reynolds numbers, one
can see low-probability instances whose gradients are of the order
of $\theta_{rms}/\eta_B$. This particular gradient value is
indicated in the lower right figure by the two vertical arrows.
\begin{figure}
\centerline{\includegraphics[angle=0,scale=0.7,draft=false]{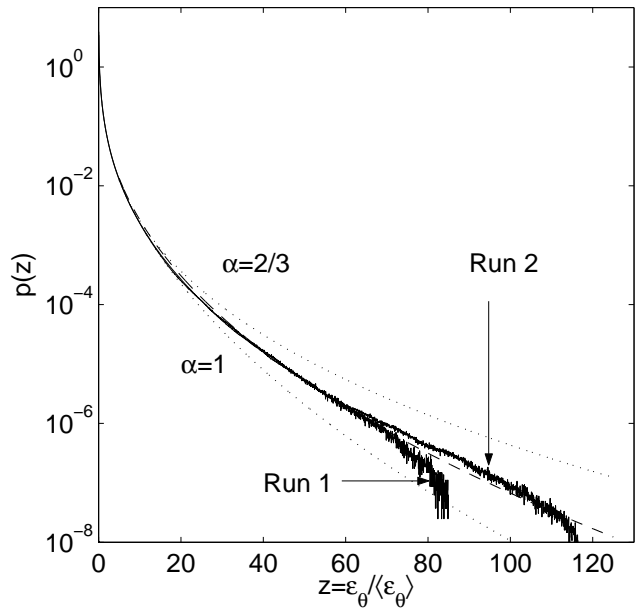}}
\caption{Log-linear plot of the probability density
function of the scalar dissipation rate, normalized to the mean
value. Data are for Run 1 and Run 2 at $Sc=32$. Fits to the data for $z\ge 10$
with the exponential term of (\ref{diffflux1}) are also plotted and the corresponding
exponents $\alpha$ are shown. The
dashed line is the optimum of a least square fit resulting in
$\alpha=0.86$.}
\end{figure}
\begin{figure}
\centerline{\includegraphics[angle=0,scale=0.7,draft=false]{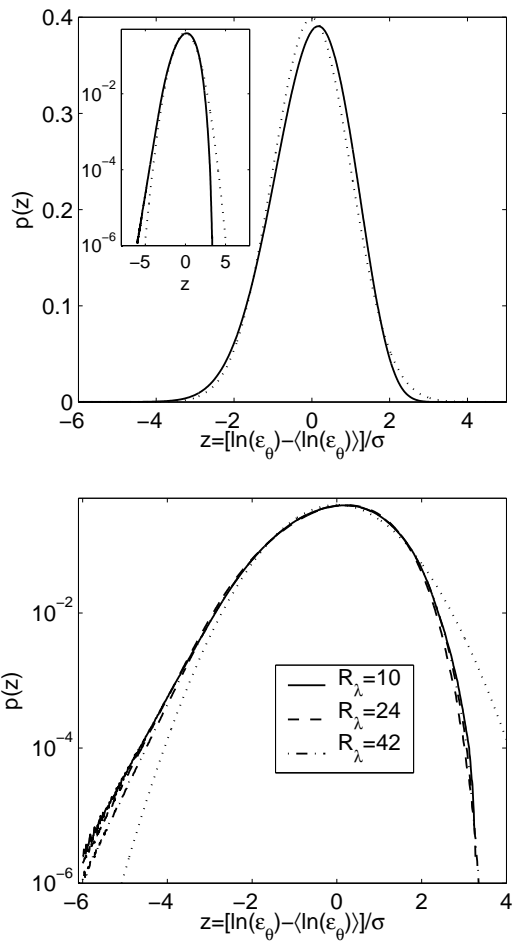}}
\caption{Upper panel: Probability density function
of $z=(\log(\epsilon_{\theta})-\langle\log(\epsilon_{\theta})\rangle)/\sigma$
with $\sigma=\sqrt{\langle(\log(\epsilon_{\theta}))^2\rangle-\langle\log(\epsilon_{\theta})\rangle^2}$.
Data for Run 2, with $Sc=32$, are compared with the corresponding
lognormal distribution having the same mean and standard
deviation. Data in each panel are plotted as a solid line and the
lognormal distribution as a dotted line. The inset shows the same
data in a log-linear plot that highlights the tails. Deviations
from lognormality are quite visible. Lower panel: Log-linear plots
of the same type of data for three different
Reynolds numbers and $Sc=32$.}
\end{figure}
\begin{figure}
\centerline{\includegraphics[angle=0,scale=0.7,draft=false]{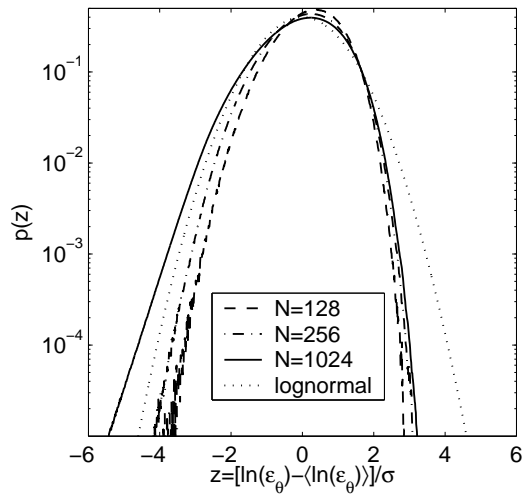}}
\caption{Log-linear plot of the PDF of the scalar
dissipation rate at $Sc=32$ for Run 2. As in Fig.~7, solid line
are the original data at a resolution of $N=1024$ and the dotted
line is for the corresponding lognormal distribution. The analysis
was repeated for the coarse-grained scalar dissipation field, and
the dashed line shows the result of coarse-graining over $8^3$
cubes, giving an effective resolution of $N=128$; the
dashed-dotted curve is for $4^3$ cubes with an
effective resolution of $N=256$.}
\end{figure}

\section{Statistics of scalar dissipation rate}
\subsection{The PDF of the scalar dissipation rate}
The scalar dissipation rate plays a central role in determining
turbulent mixing. In combustion, for example, its properties enter
fast chemistry models \cite{bilger76}, modelling of thin reactive
layers called flamelets \cite{peters86} which are embedded and
advected in the turbulent flow, and conditional-moment closures
for the evolution of the mixture fraction in non-premixed cases
\cite{klimenko99}.

With
$p(|{\bf\nabla}\theta|)\sim\exp(-(C|{\bf\nabla}\theta|^{\alpha})$
for large $|{\bf\nabla}\theta|$, as was seen in the last section,
we expect that the tails of the PDF of $\epsilon_{\theta}$ will
have stretched exponential behavior given by
\begin{equation}
p(\epsilon_{\theta}> \langle\epsilon_{\theta}\rangle)=
\frac{C_1}{\sqrt{\epsilon_{\theta}}}\,
\exp\left(-C_2\epsilon_{\theta}^{\alpha/2}\right)\,,
\label{diffflux1}
\end{equation}
where $C_1$ and $C_2$ are constants. Such statistics were derived
analytically for scalar advection in smooth and white-in-time
flows for the limit $Pe\to\infty$ and $\alpha=2/3$ was found
\cite{Falkovich98,Gamba99}. In Fig.\ 6 we show the PDF of the
scalar dissipation rate in a log-lin plot for Runs 1 and 2; Sc=32.
In addition, we fit the exponential part of the formula (\ref{diffflux1}) to the data for
the range of $\epsilon_{\theta}>10 \langle\epsilon_{\theta}\rangle$.
The cases $\alpha=1$ (pure exponential) and $\alpha=2/3$
(corresponding to the theory in Refs. 29 and 30) bound
both sets of data from below and above, respectively, with the
optimal least-square fit (dashed line) corresponding to
$\alpha=0.86$.

This form of the PDF departs clearly from log-normality.
Figure~7 (upper panel) shows the PDF of
$\epsilon_{\theta}$ for Run 2 (see Table~1). The linear scale
reveals that the lognormal approximation is not particularly good
near the core. The logarithmic scale (inset) shows that the tails depart
from lognormality as well: low dissipation values show a fatter
tail and high dissipation values fall short of the lognormal.
These departures from lognormality are robust for all Reynolds and
Schmidt numbers considered here: this is demonstrated in the lower
panel of Fig.~7 where we plot the data for $Sc=32$ at the three
Reynolds numbers $R_{\lambda}$=10, 24, and 42.

There has been much discussion in the literature about the
lognormality or otherwise of scalar dissipation, and we might cite
an early paper Sreenivasan and Antonia \cite{sreenivasan} and a
recent one by Su and Clemens \cite{su03} to illustrate experimental
results and the discussions accompanying them. Often, measured
deviations from lognormality are attributed to poor measurement
resolution or to low Reynolds numbers or to noise in the
experimental system.

In order to shed some light on how the PDFs depend on probe
resolution in an experiment, we plot in Fig.~8 the PDFs that
result when the statistics are taken over successively
coarse-grained scalar dissipation fields. The coarse-graining
scale does not produce a qualitatively different result on the
right side of the distribution ($z>0$), for which the PDF remains
sub-Gaussian for all cases. For negative $z$, there is probably a
particular coarse-graining that may collapse the data on to the
lognormal curve, but this coincidence would be accidental at best.
On the effect of Reynolds number, we can only comment that the
behavior is similar for both Reynolds numbers considered here. The
double-precision calculations throughout this work
keep the noise effects under
control. We thus conclude that, while the PDF is roughly
lognormal, the deviations from it are considerable.

\subsection{The joint statistics of scalar and scalar dissipation}
In Fig.~9 we show the joint PDF $p(\theta,\epsilon_{\theta})$
where
$p(\theta,\epsilon_{\theta})\,\mbox{d}\theta\,\mbox{d}\epsilon_{\theta}$
is the probability of having the scalar dissipation rate between
$\epsilon_{\theta}$ and
$\epsilon_{\theta}+\mbox{d}\epsilon_{\theta}$ {\em and} the scalar
fluctuation between $\theta$ and $\theta+\mbox{d}\theta$. The
figure shows the joint PDF for two Reynolds numbers and $Sc = 32$.
The upper row of Fig.~9 shows iso-contours of
$p(\theta,\epsilon_{\theta})$ which are equidistant in units of
the logarithm to base 10. The joint PDF is wider with respect to
the $\theta$-axis for the case of larger Reynolds number. This is
due to the fact that the scalar field and the dissipation field
acquire larger amplitudes with increasing $R_{\lambda}$. The lower
panels show the scalar PDF calculated by integrating
$p(\theta,\epsilon_{\theta})$ over $\epsilon_{\theta}$. We observe
an almost Gaussian shape of $p(\theta)$ for both Reynolds numbers
(as anticipated in Sec.\ III.B). Additionally, we have compared our
results with the limiting shape for a scalar field PDF that was
analytically predicted for advection in a random flow
\cite{yakhot89}. The comparison is good only for small amplitudes
of $\theta$.

A nearly Gaussian PDF for the scalar fluctuations is in agreement
with a number of other studies at $Sc\lesssim 1$, such as heated
grid turbulence \cite{mydlarski98}, isotropic turbulence with a mean
scalar gradient \cite{overholt,celani01}, a homogeneous shear flow
with heating by a constant mean temperature gradient
\cite{ferchichi}. In our simulations, the fluctuations extend to
approximately $\pm 4.5\theta_{rms}$ and the deviations in the tails
are slightly sub-Gaussian. Exponential tails of the PDF, as found by
Jayesh and Warhaft \cite{jayesh}, were not observed for our range of
parameters. It should be mentioned that they were also predicted
analytically for a white-in-time smooth Gaussian flow by Shraiman
and Siggia \cite{shraiman}. Furthermore, exponential tails have been
suggested to result from an aggregation process of elementary local
straining motion events, and close-to-exponential tails have been
observed in jets sufficiently far away from the nozzle exit
\cite{villermaux03}.

Overholt and Pope \cite{overholt} argued that the nearly Gaussian or
slightly sub-Gaussian statistics may result from the limited size of
the simulation box in relation to the integral scale. The ratio of
our box size to the integral scale of the flow, $2\pi/L$, is about 6
to 8 (see Table 1); in terms of the integral scale of the passive
scalar at the largest Schmidt number, the ratio $2\pi/L_{\theta}$
varies from 13 to 21. This is significantly larger than the
corresponding scale ratios in the experiment of Jayesh and Warhaft.
The available evidence thus suggests that the relative size of the
box size to the integral scale is not responsible for the shape of
the PDF. A plausible argument can be made that the ratio
$L/L_{\theta}$ might be a more appropriate quantity. We have
attempted to vary this ratio in the simulations, but the results are
still not conclusive.

The average of the scalar dissipation rate conditioned on the
relation $\theta({\bf x},t)=\psi$ is denoted by $\langle
\epsilon_{\theta} |\theta=\psi\rangle$. The conditional average is
defined by means of the conditional probability density function
$p(\epsilon_{\theta}|\theta)$, as
\begin{eqnarray}
\langle
\epsilon_{\theta}|\theta=\psi\rangle=\int_0^{\infty}\,\mbox{d}\epsilon_{\theta}\,
\epsilon_{\theta}\, p(\epsilon_{\theta}|\theta=\psi)\,.
\end{eqnarray}
The conditional probability density function can be expressed via
the joint probability density function of having the scalar
dissipation between $\epsilon_{\theta}$ and
$\epsilon_{\theta}+\mbox{d}\epsilon_{\theta}$ {\em and} the scalar
between $\theta$ and $\theta+\mbox{d}\theta$, so that
\begin{eqnarray}
p(\epsilon_{\theta}|\theta=\psi)=
\frac{p(\theta=\psi,\epsilon_{\theta})}{p(\theta=\psi)}\,.
\end{eqnarray}
Figure~10 shows the conditional moments for $Sc=2$ and 32. It is
well known that the conditional moment is constant \cite{obrien}
if the scalar distribution is strictly Gaussian. As can be seen in
Fig.~10 our numerical results confirm this prediction for scalar
fluctuation levels of less than about three standard deviations.
For larger scalar fluctuation levels, there is a considerable drop
in the conditional expectation. This suggests a precipitous drop
in $\epsilon_\theta$ for large values of $\theta$, in contrast,
say, to the behavior of the energy dissipation field $\epsilon$
with respect to the velocity fluctuations \cite{krs-sd}.

Experiments in inhomogeneous flows \cite{kailas} show that the
constancy of the conditional moment holds roughly on the
centerline of turbulent jets and wakes. The behavior off-center is
quite complex, and the boundary layer flows are intrinsically
different.
\begin{figure}
\centerline{\includegraphics[angle=0,scale=0.8,draft=false]{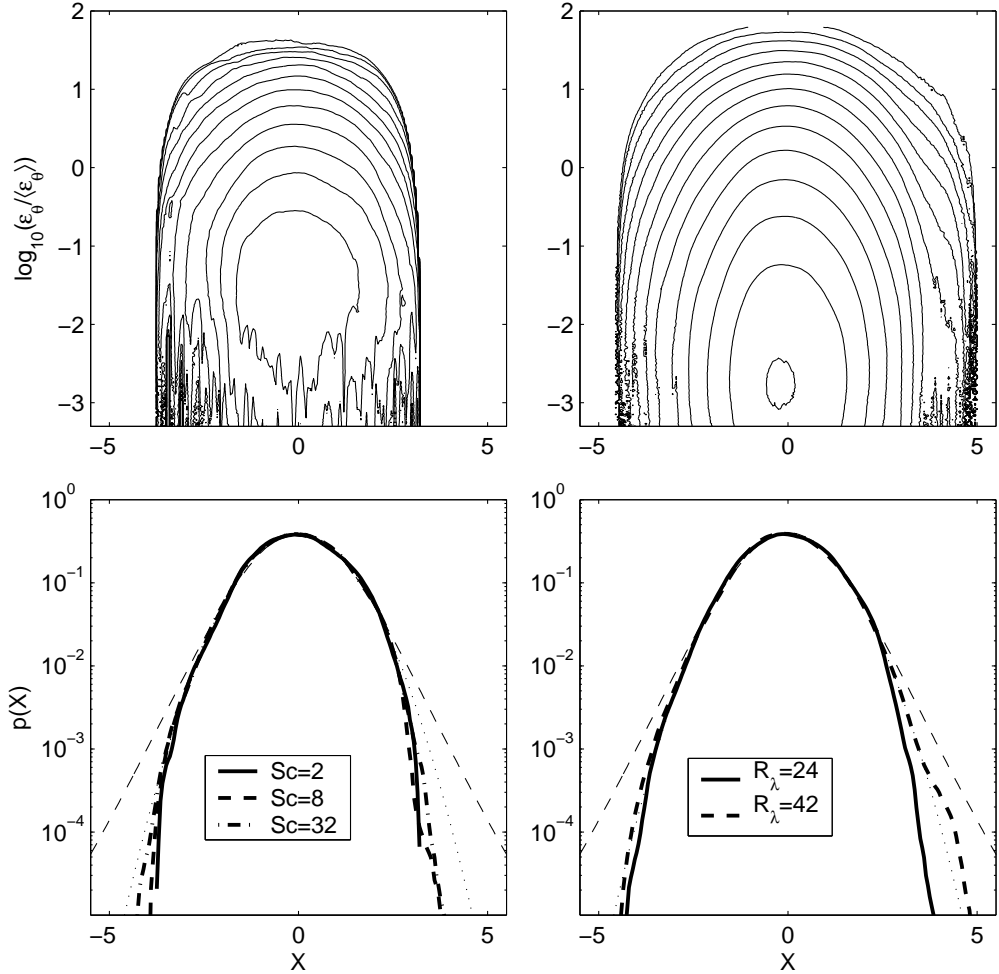}}
\caption{Upper row: Joint probability density function
$p(\theta/\theta_{rms},\epsilon_{\theta}/\langle\epsilon_{\theta}\rangle)$
for the runs at $R_{\lambda}=24$ (left) and  $R_{\lambda}=42$
(right), both for $Sc=32$. The iso-contour levels are equidistant
in units of the logarithm to base 10, varying from $-6$ to 0.5 in
steps of 0.5. The joint densities for large negative values of
log$(\epsilon_\theta)$ (i.e., scalar dissipation rates smaller
than $10^{-3}\langle\epsilon_\theta\rangle$) are truncated for reasons of poor
convergence. Lower row: scalar PDF for $R_{\lambda}=24$ and $Sc$
values indicated in the legend (left). The thin dashed line is a
theoretical prediction for the scalar PDF,
$p(X)=C/(1+kX^2)^{1+1/2k}$, for advection in a random flow with
fit parameters $C=0.4$, $k=0.08$, $X$ being $\theta/\theta_{rms}$
\cite{yakhot89}. The thin dotted line is a Gaussian distribution.
Scalar PDF at $Sc=32$ for two different Reynolds numbers as
indicated in the legend are shown in the right panel. Both thin
lines are as in the lower left panel.}
\end{figure}
\begin{figure}
\centerline{\includegraphics[angle=0,scale=0.7,draft=false]{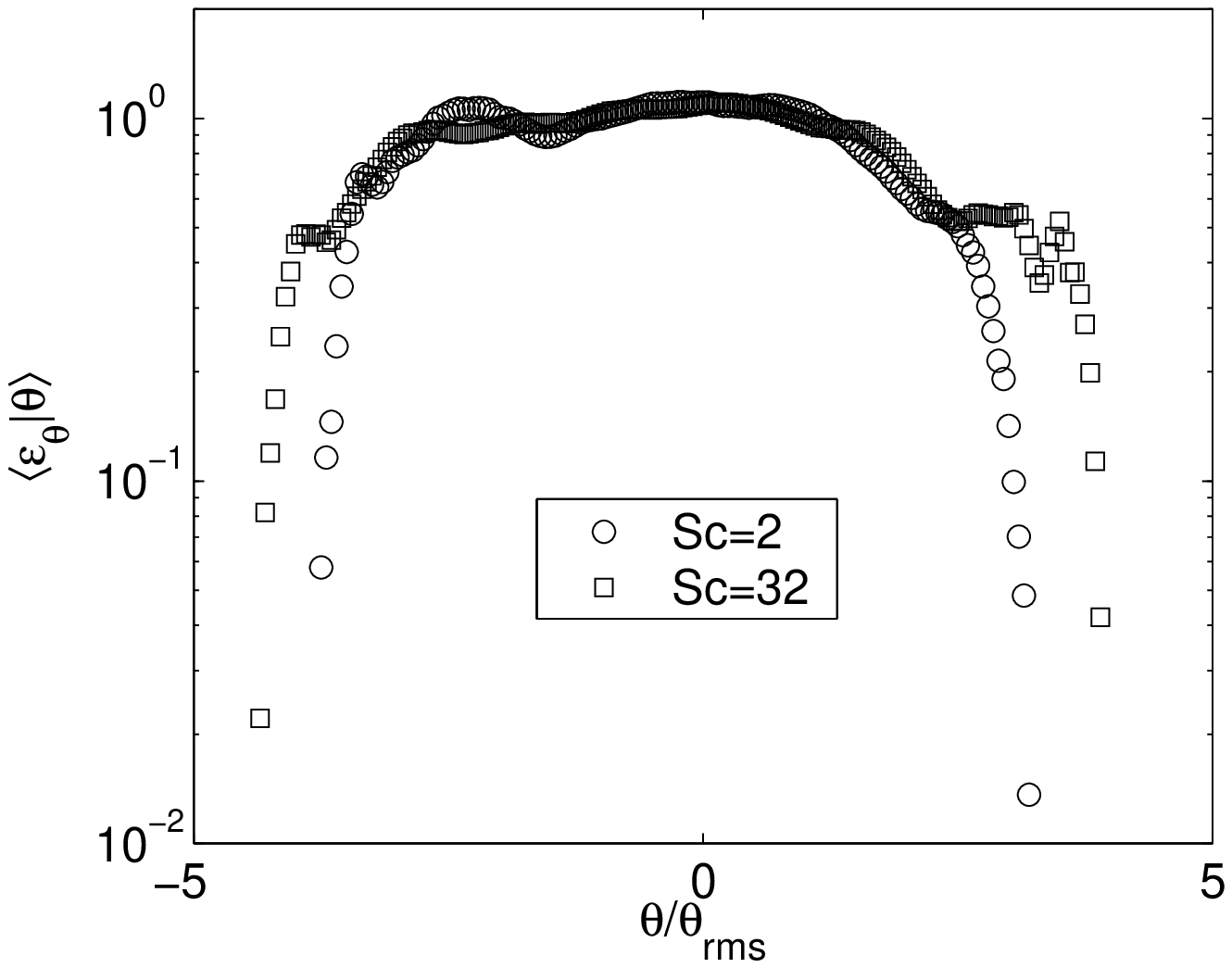}}
\caption{Mean conditional moment of the scalar
dissipation rate for Schmidt numbers $Sc=2$ and 32. Data are for
Run 2 in Table 1.}
\end{figure}

\section{Summary and conclusions}

We have studied numerically the dynamics of passive scalars
advected in homogeneous isotropic turbulence for three Schmidt
numbers $Sc>1$ and three Reynolds numbers, $R_{\lambda}$. These
Reynolds and Schmidt numbers are moderate but comparable to those
of many laboratory experiments on reacting flows. Usually, in such
flows, neither the inertial range nor the viscous-convective range
is well developed. The most important range of scales then belong
to the so-called intermediate viscous range, which essentially
pertain to the crossover between the inertial and viscous ranges
of scales \cite{frisch91,schu05}. In this range, rough velocity
filaments reach down into the viscous range below the Kolmogorov
scale $\eta$ and stir the scalar. As reported in Ref. 16,
the intermediate viscous range gets more extended with increasing
Reynolds number. We have paid particular attention to the grid
resolution of the simulations, which exceeds the nominal value by
a factor of four, to ensure that the high intensity events of the
scalars and their gradient fields are resolved well.

The first quantity considered is the area-to-volume ratio of
isoscalar surfaces because of its fundamental relevance to
reacting flows. For the Reynolds and Schmidt numbers considered
here, the fractal scaling of isoscalar level sets is not obtained
and so cannot be used to estimate the area-to-volume ratio. This
ratio, computed here directly, increases with an increase in the
passive scalar isolevel, the Schmidt number, and the Reynolds
number, respectively. In detail, these increasing trends depend on
the statistics of the scalar itself, which is found here to be
nearly Gaussian.

We have defined a so-called transition scale $\langle \ell_X \rangle$ that becomes
smaller with increasing surface-to-volume ratio of the isosurface
corresponding to the scalar magnitude $X$ (normalized by the
root-mean-square value). Quite understandably, the isosurface
becomes "rougher" and more disconnected with increasing $Sc$ and increasing Reynolds
number, but is not dependent on $X$ itself---except when $X$ is
large. The flat behavior of the scale $\langle\ell_X\rangle$ for
$X\lesssim 2$ can be considered a geometric confirmation of the
nearly Gaussian probability density function $p(X)$ of the scalar.
For large $X$, there is a tendency for $\langle \ell_X \rangle$ to
decrease with $X$, this being attributable to the sub-Gaussian
behavior of $p(X)$ in the tail region. The flux across the
interface is a strongly fluctuating quantity from one differential
area element to another, and its distribution is closely
exponential. The fluctuations increase rapidly with increase in
$Sc$ as well as $R_\lambda$. Though with low probability,
gradients of the order of magnitude $\theta_{rms}/\eta_B$ do occur
in the scalar field. This clearly suggests the existence of sharp
fronts accompanied by large jumps in the scalar concentration.

The scalar dissipation rate departs from lognormality more than
previous measurements suggest. The reasons for close
correspondence in previous data are not entirely clear, but it
must be pointed out that very few data sets exist in which all
three scalar gradient components that enter the scalar dissipation
rate definition (see (Eq.~\ref{scalardiss})) were measured with
adequate resolution and without the use of Taylor's hypothesis. To
consider if the closeness to lognormality observed previously was
caused by poor resolution in measurements, we coarse-grained the
data and replotted them. The coarse-grained data also display
significant departures from lognormality. We thus conclude that
the reasons for observed departures are more complex; they may
also emphasize the lack of strict universality in the dissipation
statistics.

The high spectral resolution in the present DNS allowed the
evaluation of a conditional moment analysis as well. We have been
able to confirm theoretical predictions that follow from nearly
Gaussian fluctuations of the passive scalar field for $\theta < 3
\theta_{rms}$, say.

In summary, we have quantified various properties of the scalar
isosurfaces for conditions that are typical of many laboratory
experiments in reacting flows. Beside the connections between
statistical and geometrical properties of scalars studied here, it
is interesting to shed more light on the formation of
characteristic scalar structures with respect to time, and to
relate them to the underlying local flow properties for different
Reynolds numbers. This can be a useful way of incorporating
Reynolds number effects on mixing in the Batchelor regime and is
part of the future work.

\acknowledgements 
The computations were carried out on
up to 256 IBM Power4 CPUs of the J\"ulich Multiprocessor (JUMP) machine at
the John von Neumann-Institute for Computing of the Research
Centre J\"ulich (Germany). We acknowledge their steady support,
and the support by the Deutsche Forschungsgemeinschaft (to JS) and
by the US National Science Foundation (to both JS and KRS).
JS wants to thank the Institute for Pure and Applied Mathematics at 
UCLA for hospitality during the Multiscale Geometric Analysis program
where parts of this work were done. We
also thank R.W. Bilger, W.J.A. Dahm, J. Davoudi, J.A. Domaradzki,
B. Eckhardt, S.B. Pope and P.K. Yeung for useful discussions.


\begin{thebibliography}{99}
\bibitem{prasad90}
R.R. Prasad and K.R. Sreenivasan, ``Quantitative three-dimensional
imaging and the structure of passive scalar fields in fully
turbulent flows," J.\ Fluid Mech.\ {\bf 216}, 1 (1990).

\bibitem{dahm96}
K.A. Buch, Jr. and W.J.A. Dahm, ``Experimental study of the
fine-scale structure of conserved scalar mixing in turbulent shear
flows. Part 1. $Sc \gg 1$," J.\ Fluid Mech.\ {\bf 317}, 21 (1996).

\bibitem{Dimotakis96}
H.J. Catrakis and P.E. Dimotakis, ``Scale distributions and fractal dimensions
in turbulence," Phys. Rev. Lett. {\bf 77}, 3795 (1996).

\bibitem{villermaux99}
E. Villermaux and C. Innocenti, ``On the geometry of turbulent
mixing," J.\ Fluid Mech.\ {\bf 393}, 123 (1999).

\bibitem{catrakis02}
H.J. Catrakis, R.C. Aguirre and J. Ruiz-Plancarte,``Area-volume
properties of fluid interfaces in turbulence: scale-local
self-similarity and cumulative scale dependence," J.\ Fluid Mech.\
{\bf 462}, 245 (2002).

\bibitem{su03}
L.K. Su and N.T. Clemens,``The structure of fine-scale scalar mixing
in gas-phase planar turbulent jets," J.\ Fluid Mech.\ {\bf 488}, 1
(2003).

\bibitem{long94}
S.H. St{\aa}rner, R.W. Bilger, K.M. Lyons, J.H. Frank and M.B.
Long, ``Conserved scalar measurements in turbulent-diffusion
flames by a Raman and Rayleigh ribbon imaging method," Combust.\
Flame {\bf 99}, 347 (1994).

\bibitem{karpetis02}
A.N. Karpetis and R.S. Barlow, ``Measurements of scalar
dissipation in a turbulent piloted methane/air jet flame," Proc.\
Combust.\ Inst.\ {\bf 29}, 1929 (2002).

\bibitem{batchelor} G.K. Batchelor, ``Small-scale variation of
convected quantities like temperature in a turbulent fluid. Part
1. General discussion and the case of small conductivity," J.\
Fluid Mech.\ {\bf 5}, 113 (1959).

\bibitem{bilger04} R.W.~Bilger, ``Some aspects of scalar dissipation,"
Flow Turb.\ Combust. {\bf 72}, 93 (2004).

\bibitem{yeung01}
P. Vedula, P.K. Yeung and R.O. Fox, ``Dynamics of scalar
dissipation in isotropic turbulence: a numerical and modelling
study," J.\ Fluid Mech.\ {\bf 433}, 29 (2001).

\bibitem{yeung02}
P.K. Yeung, S. Xu and K.R. Sreenivasan, ``Schmidt number effects
on turbulent transport with uniform mean scalar gradient," Phys.\
Fluids {\bf 14}, 4178 (2002).

\bibitem{brethouwer03}
G. Brethouwer, J.C.R. Hunt and F.T.M. Nieuwstadt,
``Micro-structure and Lagrangian statistics of the scalar field
with a mean gradient in isotropic turbulence," J. Fluid Mech. {\bf
474}, 193 (2003).

\bibitem{schu03}
J. Schumacher and K.R. Sreenivasan, ``Geometric features of the
mixing of passive scalars at high Schmidt numbers," Phys.\ Rev.\
Lett.\ {\bf 91}, 174501 (2003).

\bibitem{watanabe04}
T. Watanabe and T. Gotoh, ``Statistics of passive scalar in
homogeneous turbulence," New J. Phys. {\bf 6}, 40 (2004).


\bibitem{schu05} J. Schumacher, K.R. Sreenivasan and P.K. Yeung,
``Very fine structures in scalar mixing," J.\ Fluid
Mech.\ {\bf 531}, 113 (2005).

\bibitem{schu04} J. Schumacher, ``Relation between shear parameter
and Reynolds number in statistically stationary shear flows,"
Phys.\ Fluids {\bf 16}, 3094 (2004).

\bibitem{Ott} E. Ott, {\em Chaos in dynamical systems},
Cambridge University Press, Cambridge, 2002.

\bibitem{sreeni} K.R. Sreenivasan, R. Ramshankar and C. Meneveau,
``Mixing, entrainment, and fractal dimension of interfaces in
turbulent flows," Proc.\ Roy.\ Soc.\ Lond. {\bf 421A}, 79 (1989).

\bibitem{Falconer} K.J. Falconer, {\em The geometry of fractal sets},
Cambridge University Press, Cambridge, 1985.

\bibitem{frederiksen96}
R.D. Frederiksen, W.J.A. Dahm, and D.R. Dowling, ``Experimental
assesment of fractal scale-similarity in turbulent flows. Part 1.
One-dimensional intersections," J.\ Fluid Mech. {\bf 327}, 35
(1996).

\bibitem{pope88} S.B. Pope, ``The evolution of surfaces in turbulence,"
Int.\ J.\ Eng.\ Sci.\ {\bf 26}, 445 (1988).

\bibitem{aguirre04}
R. C. Aguirre and H. J. Catrakis, ``Interfacial-fluid dynamics and
the mixing efficiency of turbulent flows," Phys.\ Fluids {\bf 16},
4746 (2004).

\bibitem{verwisch98} L.
Verwisch and T. Poinsot, ``Direct numerical simulation of
non-premixed turbulent flames," Annu.\ Rev.\ Fluid Mech. {\bf 30},
655 (1998).

\bibitem{Kollmann1997} W. Kollmann and J.J. Lienau,
``Computation of topological and geometric properties of
isosurfaces I: theory and numerical method," Int.\ J.\ Comp.\
Fluid Dyn.\ {\bf 8}, 83 (1997).

\bibitem{bilger76}
R.W.~Bilger, ``Turbulent jet diffusion flames," Progr.\ Energy
Combust.\ Sci.\ {\bf 1}, 87 (1976).

\bibitem{peters86}
N. Peters, ``Laminar flamelet concepts in turbulent combustion,"
Proc.\ Combust.\ Inst.\ {\bf 21}, 1231 (1986).

\bibitem{klimenko99}
A.Y. Klimenko and R.W. Bilger, ``Conditional moment closure for
turbulent combustion," Progr.\ Energy Combust.\ Sci.\ {\bf 25},
595 (1999).

\bibitem{Falkovich98}
M. Chertkov, G. Falkovich and I. Kolokolov, ``Intermittent
dissipation of a passive scalar in turbulence," Phys.\ Rev.\ Lett.
{\bf 80}, 2121 (1998).

\bibitem{Gamba99}
A. Gamba and I. Kolokolov, ``Dissipation statistics of a passive
scalar in a multi-dimensional smooth flow,"  J.\ Stat.\ Phys. {\bf
94}, 759 (1999).

\bibitem{sreenivasan} K.R. Sreenivasan, R.A. Antonia and H.Q.
Dahn, ``Temperature dissipation fluctuations in a turbulent
boundary layer," Phys.~Fluids {\bf 20}, 1238 (1977).

\bibitem{yakhot89}
Y.G. Sinai and V. Yakhot, ``Limiting probability distributions of
a passive scalar in a random velocity field," Phys.\ Rev.\ Lett.\
{\bf 63}, 1962 (1989).

\bibitem{mydlarski98}
L. Mydlarski and Z. Warhaft, ``Passive scalar statistics in
high-P\'{e}clet number grid turbulence," J.\ Fluid Mech. {\bf
358}, 135 (1998).

\bibitem{overholt}
M. R. Overholt and S. B. Pope, ``Direct numerical simulation of a
passive scalar with imposed mean gradient in isotropic
turbulence,"  Phys.\ Fluids {\bf 8}, 3128 (1996).

\bibitem{celani01}
A. Celani, A. Lanotte, A. Mazzino and M. Vergassola, ``Fronts in
passive scalar turbulence," Phys.\ Fluids {\bf 13}, 1768 (2001).

\bibitem{ferchichi}
M. Ferchichi and S. Tavoularis, ``Scalar probability density
function and fine structure in uniformly sheared turbulence," J.\
Fluid Mech. {\bf 461}, 155 (2002).

\bibitem{jayesh}
Jayesh and Z. Warhaft, ``Probability distribution of a passive
scalar in grid-generated turbulence," Phys.\ Rev.\ Lett. {\bf 67},
3503 (1991).

\bibitem{shraiman} B. I. Shraiman and E. D. Siggia, ``Lagrangian path integral
and  fluctuations in random flow," Phys.\ Rev.\ E {\bf 49}, 2912
(1994).

\bibitem{villermaux03}
E. Villermaux and J. Duplat, ``Mixing as an aggragation process,"
Phys.\ Rev.\ Lett. {\bf 91}, 184501 (2003).

\bibitem{obrien} F. Gao and E.E. O'Brien, ``Joint probability density
function of a scalar and its gradient in isotropic turbulence,"
Phys.\ Fluids A {\bf 3}, 1625 (1991).

\bibitem{krs-sd} K.R. Sreenivasan and B. Dhruva,
``Is there scaling in high-Reynolds-number turbulence?"
Prog.~Theo.~Phys. {\bf 130}, 103 (1998).

\bibitem{kailas} P. Kailasnath, K.R. Sreenivasan and J.R. Saylor,
``Conditional scalar dissipation rates in turbulent wakes, jets
and boundary layers," Phys.\ Fluids {\bf A5}, 3207 (1993).

\bibitem{frisch91} U. Frisch and M. Vergassola, ``A prediction of the
multifractal model: the intermediate dissipation range,"
Europhys.\ Lett. {\bf 14}, 439 (1991).

\bibitem{yeung05} D.A. Donzis, P.K. Yeung, and K.R.Sreenivasan,
``Scalar dissipation rate and dissipative anomaly in isotropic turbulence,"
J. Fluid Mech. {\bf 532}, 199 (2005).
\end{thebibliography}
\end{document}